\documentclass[11pt]{article}
\usepackage[utf8]{inputenc}
\usepackage{amsmath}
\usepackage{amssymb}
\usepackage[margin=1in]{geometry}
\usepackage{hyperref}
\hypersetup{breaklinks=true}
\usepackage{cleveref}
\usepackage{setspace}
\usepackage{titlesec}
\usepackage[normalem]{ulem}

\usepackage{tikz}
\usetikzlibrary{arrows.meta}
\usepackage[style=numeric,citestyle=numeric-comp,backend=biber,giveninits=true,backref=true,maxnames=10,sortcites=false,sorting=nyt]{biblatex}
\DeclareNameFormat{author}{%
  \ifthenelse{\value{listcount}=1}
    {%
      \ifdefvoid{\namepartprefix}{}{\namepartprefix\addspace}%
      \namepartfamily\addcomma\addspace\namepartgiveni
    }
    {\usebibmacro{name:given-family}
       {\namepartfamily}{\namepartgiveni}{\namepartprefix}{\namepartsuffix}}%
  \usebibmacro{name:andothers}}

\DeclareNameFormat{editor}{%
  \ifthenelse{\value{listcount}=1}
    {%
      \ifdefvoid{\namepartprefix}{}{\namepartprefix\addspace}%
      \namepartfamily\addcomma\addspace\namepartgiveni
    }
    {\usebibmacro{name:given-family}
       {\namepartfamily}{\namepartgiveni}{\namepartprefix}{\namepartsuffix}}%
  \usebibmacro{name:andothers}}
\newbibmacro{string+doi}[1]{%
  \iffieldundef{doi}{#1}{\href{http://dx.doi.org/\thefield{doi}}{#1}}}
\DeclareFieldFormat{title}{\usebibmacro{string+doi}{\mkbibemph{#1}}}
\DeclareFieldFormat[article]{title}{\usebibmacro{string+doi}{\mkbibquote{#1}}}
\renewbibmacro{in:}{}
\renewbibmacro*{volume+number+eid}{%
  \printfield[bold]{volume}
  \printfield[parens]{number}%
  \setunit{\addcomma\space}%
  \printfield{eid}}

\counterwithout{figure}{section}
\renewcommand{\thesection}{\arabic{section}}
\titleformat{\section}
  {\normalfont\bfseries\centering}{\thesection.}{0.3em}{}

\makeatother

\begin{document}
\title{{\Huge QBism on Locality and Nonlocality\\
}\medskip{}
}
\author{John B. DeBrota$^1$ \and Christopher A. Fuchs$^2$ \and R\"udiger Schack$^3$\thanks{We thank all those who persistently questioned QBism's locality over the years or with whom we have had stimulating discussions on the topic. A partial list includes Eric Cavalcanti, Gino Elia, Joe Henson, Ruth Kastner, Christian List, Alyssa Ney, Govind Sidhardh, Jacques Pienaar, Blake Stacey, and Matt Weiss. Special thanks go to science journalist George Musser for issuing the \emph{Musser Challenge}: ``I cannot write about how QBism evades Bell's argument until I understand it.'' Without his prod, this paper might have slept in the crypt another 13 years before seeing sunlight.}}
\date{%
    $^1$\small Munich Center for Mathematical Philosophy, LMU Munich\\%
    $^2$Department of Physics, University of Massachusetts Boston\\
    $^3$Department of Mathematics, Royal Holloway University of London\\[2ex]%
}\vspace{-2cm}

\maketitle
\hrule
\vspace{1em}
{\setstretch{1.0}
\noindent 
Recently Pienaar (2026), building on work of Cavalcanti (2021), has shown that QBism cannot always assume distinct observers' quantum-measurement outcomes---say, of Wigner and his friend---are embedded in a single spacetime. This follows from QBism's rejection of the `Absoluteness of Observed Events' assumption in the Bong et al.\ no-go theorem. Thus, QBism has no choice but to treat the notion of spacetime every bit as personalistic as it treats quantum states and quantum measurement outcomes. In a way, this is not a surprise to QBists, as they have taken the notion of `personalist spacetimes' to be the \emph{ansatz} most compatible with their other views since at least 2009. But it does enjoin us to finally make crystal clear the sense in which QBism is a purely local interpretation of quantum mechanics despite this new theorem and despite quantum theory's age-old violation of Bell's inequalities. With the extra clarity we also hope to poise QBism for a distinctly new way to approach issues at the interface of quantum theory and gravity.

\par
}
\vspace{1em}
\hrule
\vspace{1em}
\section{Introduction}

Locality is a cherished ideal in physics. It is always considered an advantage when a theory can be said to be local. When locality is questioned, the whole theory is called into question. More specifically, when it is supposed that causal notions play a role in physics, causes are traditionally taken to be spatially localized events that are constrained in two independent ways: first, they ``do not act where they are not'' and, second, causal influences ``do not propagate infinitely fast''~\cite{Frisch2025}. Newton famously considered the violation of locality in his theory of gravitation to be an unacceptable shortcoming, indicating that it cannot be the final story \cite{Berkovitz2007}. In turn, one of the crowning achievements of Einstein's general relativity was the removal of this shortcoming. Classical electrodynamics can be formulated as a nonlocal theory involving particles without fields (say, by the Wheeler-Feynman scheme \cite{Wheeler1945,Wheeler1949}), but the availability of the standard treatment overwhelmingly secures the theory as a point in favor of the local nature of reality.

Quantum mechanics has been the most persistent challenger to the physical norm of locality. While it is recognized that rejecting locality is not the only acceptable reaction to challenges like Bell's theorem \cite{Bell1964}, a number of influential philosophers of science strongly favor this escape route. The most polemic analysis comes from Maudlin~\cite{Maudlin2014}, who argues that rejecting locality is strictly the only option. Others, like Myrvold~\cite{Myrvold2016}, argue that a rejection of locality is the most defensible option, but admit that a rejection of `measurement independence'---that is, the demand that measurement settings be statistically independent of the system in question---remains a live option, even if theoretically unattractive.\footnote{Interpretations which claim to preserve both `locality' and `realism' by relaxing `measurement independence' include superdeterministic theories and retrocausal theories. Although preserving the ideal of locality, these come with steep costs of their own, for instance, by calling into question the sense in which experimenters can actually choose their own settings and introducing the necessary fine-tuning drawn attention to by Wood and Spekkens~\cite{Wood2015}.}  
However, in addition to subverting the ideal of locality, this move also likely requires making unappealing modeling choices. For example, any explicitly nonlocal proposal has to address how it is that there are instantaneous influences yet no possibility of `signaling', that is, the transmission of information faster than the speed of light (or any speed at all, for that matter). 

It is hard to find, among those philosophers who champion nonlocality, anyone who believes the third escape route---rejecting Bell's `realism'---is even coherent. Nevertheless, a majority of physicists who self-identify as subscribers to the Copenhagen interpretation~\cite{Gibney2025}, reject Bell's `realism' and still usually claim to maintain locality in some fashion. Whereas this path avoids the need to explicitly reject locality, the sense in which one can positively affirm locality has to be addressed on a case-by-case basis. Perhaps partially in light of this vagueness, there is a cultural sense that quantum mechanics has identified an essential nonlocality in nature. Indeed, the robust experimental verification of Bell inequality violations \cite{Giustina2015} is bandied as a confirmation of Einstein's reviled ``spooky action at a distance'' in popular media \cite{Musser2015} and research into this feature of quantum mechanics sometimes goes under the broad heading of `quantum nonlocality' even when a researcher might not believe it to be exactly about that.\footnote{An amusing anecdote comes from Valerio Scarani in a private announcement of his 2019 book {\sl Bell Nonlocality}~\cite{Scarani2019}.  He wrote to his distribution list, ``[F]or the few of you who will feel itchy about the title, feel free to say that you don't like it, but don't attribute to me statements like `nature is nonlocal': it's not written anywhere, I left it for anyone to make their own call.''}

Our concern in this paper is with QBism \cite{Fuchs2023,Fuchs2013,Fuchs2018,Mermin2019}. Most radical among the `Copenhagenish' and `perspectival' interpretations \cite{Schmid2025}, QBism summarily rejects Bell's realism alongside many other traditional ontological commitments. Yet, for all that it abandons of these traditions, QBism claims to be local. In fact, QBists consider locality to be of the utmost importance. Nonetheless, our experience suggests that QBism's largely apophatic stance (i.e., saying what reality \emph{is not}) has led some to believe that it lacks the means to positively assert it is local. And the worries might be deeper still: Pienaar \cite{Pienaar2026}, building on work by Cavalcanti \cite{Cavalcanti2021}, recently proved that in interpretations like QBism that reject a thesis known as the Absoluteness of Observed Events (AOE) \cite{Bong2020,Schmid2023}, one cannot assume that the outcomes of various observers are embedded in a single spacetime. If this is so, shouldn't the QBists admit that it's not really appropriate to call QBism `local' after all?  The ultimate gotcha moment?

Long before Pienaar's result, however, QBists were already asserting that different observers do not `inhabit' a single spacetime \cite{Fuchs2010,Fuchs2012,Fuchs2014,Mermin2013,Mermin2014}.\footnote{As well, a conversation with Eric Cavalcanti in 2007--2008 must have played an influential role in saying things this way (as it was a thread that goes back to 2005 in QBism~\cite{Fuchs2026}).  See~\cite[p.\ 1578]{Fuchs2014b} where Cavalcanti argues that in a Wigner's friend scenario, QBism would inevitably lead to the conclusion ``[the friend's measurement] event is in a sense not in the same space-time as you are now, if by definition every event in a space-time have \emph{some} causal relation with every other event.'' Though, he followed that by writing, ``The only way out of this conclusion, it seems to me, is to accept a hidden-variable interpretation in which there \emph{are} objective matters of fact about events, we just don't know what they are, or extreme solipsism.  But I would agree with you if you say that the alternative above sounds much more exciting.''}  Pienaar notes as well that, for QBists, spacetime ``should not be understood as an objective description of reality independently of the agent's activity'' \cite[p.\ 2]{Pienaar2026}. So, what is the QBist understanding of locality after all? How does a QBist account for quantum correlations without invoking what amounts to nonlocality? Furthermore, is it even possible for a physical theory to be nonlocal within the framework of QBism? These are the subjects of our paper.

\section{QBism and claims of locality}

QBism dissolves the standard quantum paradoxes by pulling at a personalist Bayesian thread in quantum information theory and following it to some radical metaphysical conclusions. In brief, rather than a direct description of reality, the quantum formalism becomes \emph{a decision theory shaped by reality} that can be adopted by any agent who wishes to better manage the consequences of her actions~\cite{Fuchs2023}. These consequences, in turn, are understood to be personal experiences for the agent who took the action. All of the objects of quantum theory, including quantum states, channels, and measurements, are personal judgments of the user of the theory~\cite{DeBrota2024}, and ultimately refer to her beliefs about her possible experiences. While individual objects are relative to a reasoning subject, the structure of the theory as a whole is invariant, to be adopted by anyone in order to detect inconsistencies in their beliefs.

In their broader project, QBists wish to identify ontological lessons consistent with this reading of the theory. Although their ontological proposals are still in early development, QBists call for radical departures from the traditional scientific worldview, for example, by rejecting the notion of a `view from nowhere'~\cite{Nagel1986}, by suggesting that the consequences of measurements be thought of as genuine moments of creation \cite{Gefter2024}, and by invoking analogies to William James's `pure experience' \cite{Fuchs2017b,Fuchs2023} and Merleau-Ponty's `flesh'~\cite{Bitbol2020,Bitbol2023,Bitbol2026,Schack2023}. To give the direction sought after a name, Fuchs~\cite{Fuchs2017} has proposed the term \emph{participatory realism}.

Despite being radical in other ways, the QBists have always insisted that quantum theory be understood in local terms. This is no small matter. Although perhaps difficult for its critics to believe, QBism's radical moves were not made lightly. Each stance was born from years of intense struggle for consistency. But so foundational is the conception of local action for QBism that it has never been seriously doubted. What would it mean for an agent to be autonomous or embodied without it?  Fuchs \cite{Fuchs2010} also quotes approvingly of Einstein's intuition that, ``[t]he complete suspension of this basic principle would make impossible the idea of the existence of (quasi-) closed systems and, thereby, the establishment of empirically testable laws in the sense familiar to us.''\footnote{Gino Elia points out that the deeper issue for QBism is not completely to do with localization in space, as it might have been with Einstein, but rather with the individuation of systems, or more to the point, a way for an agent to demarcate when two things are assumed to have a ``mutually independent existence.''} 

Arguments that QBism is local have taken several forms over the years. However, most are more specifically arguments against conclusions of nonlocality drawn from non-QBist ways of thinking about quantum theory.  That is to say, they were more arguments for ``not nonlocality'' than ``locality'' itself.

Take the Einstein-Podolsky-Rosen (EPR) thought experiment~\cite{Einstein1935} and Bell's additions to it. In \cite{Fuchs2010}, it was argued that one should reject the probability-1 provision of the EPR criterion of reality simply on personalist Bayesian grounds. Probabilities do not obtain different metaphysical significance when someone assigns a probability of zero or one, so even with a ``no disturbance'' clause as EPR used based on an assumption of locality, one should not assume the preexistence of an outcome just because an observer is utterly confident (probability-1) of what she will find. Consequently, if one rejects the criterion of reality in the EPR argument, one is not required to reject locality in the ways argued for by Maudlin \cite{Maudlin2014} and Norsen~\cite{Norsen2006}. See Section 2.7 of \cite{Fuchs2023} for further discussion of this point. 

Another argument is that the mathematics behind the so-called demonstrations of nonlocality would hold even if one were talking about a single localized system whose operator space ${\cal L}({\cal H})$ is decomposed into two commuting sub-algebras such that formally a tensor product is introduced.  In other words, so long as the dimension of $\cal H$ is not prime, it is always possible to write ${\cal H}={\cal H}_A\otimes {\cal H}_B$, which gives a way to think of a single system as a bipartite one.  In such cases, one would never be compelled to use Bell's reasoning for calculating one's probabilities.  Fuchs and Stacey \cite{Fuchs2025} walk through this argument in detail, but the idea is already recorded in the first textbook ever on quantum mechanics. Hermann Weyl (1928) writes: ``The kinematically independent parts into which a system can be resolved need not be spatially separated, nor need they even refer to different particles'' (\cite[p.\ 93]{Weyl1950}). All that quantum theory's violation of Bell's inequalities indicates is that if one feels compelled to use a different method of calculating probabilities than quantum theory's own, one will get a different answer.  In fact, there is a general lesson here:  In QBism, writing a tensor product does not imply an independent existence of separate things.  Such is the case in the example here, and such is the case in a QBists' translation of Bell's scenario into first-person language, as we will see in Section~\ref{AllThreeOfThem}.

A final kind of argument derives from the observation that the quantum formalism may be considered an addition to classical probability theory, and in the context of updating personal probability assignments, there is no physical nonlocality. DeBrota and Stacey, for instance, write: 
\begin{quote} 
``Compare this with classical electromagnetism: In that subject, if we could toggle a quantity at a distance but only in ways that could not effect a transmission of information, we'd have no hesitation in calling that quantity unphysical---an artifact, we'd say, of choosing a gauge that does not respect relativistic causality.'' (\cite[p.\ 5]{DeBrota2018FAQBism})
\end{quote}
Years before QBism was even QBism~\cite{Stacey2019}, Fuchs~\cite{Fuchs2002} observed that when a measurement of one subsystem produces an update on another, the operation on the other involves no specifically quantum peculiarity. Namely, performing a measurement on one side of an EPR pair merely enacts an update of the far-away system's quantum state according to Bayes' rule, \emph{not} the usual von Neumann or L\"uders collapse rules.  This is exactly what one would expect if measuring one side causes no physical disturbance to the other: The state update corresponds to simply having learned something. As Fuchs noted in a 2003 lecture,  
\begin{quote}
``Looking at quantum collapse this way turns the usual debate in quantum foundations on its head:  only local state changes look to be a mystery.  State changes at a distance (as after a measurement on one half of an EPR pair) are completely innocent---they simply correspond to applications of Bayes' rule itself \ldots [T]hat is, collapse-at-a-distance is nothing more than the usual method of updating one's information after gathering data.'' (\cite[p.\ 752]{Fuchs2014b})
\end{quote}

These remain good arguments, but an argument for why QBism is explicitly local---not simply just ``not nonlocal''---would be preferred if possible. Fuchs, Mermin, and Schack took a stab at it in 2014:
\begin{quote} 
``Quantum correlations, by their very nature, refer only to time-like separated events: the acquisition of experiences by any single agent. Quantum mechanics, in the QBist interpretation, cannot assign correlations, spooky or otherwise, to space-like separated events, since they cannot be experienced by any single agent. Quantum mechanics is thus explicitly local in the QBist interpretation.'' (\cite[p.\ 750\textendash751]{Fuchs2014})
\end{quote} 
It is certainly true that an agent's experiences always come to her along her time-like worldline, and this fact establishes a kind of irrevocable locality to the interpretation. However, thinking this is the only meaningful positive sense in which QBism is local is a critical oversight.\footnote{Credit is due to Joe Henson, who seems to have first drawn attention to this point in a 2015 talk. Henson's argument was later discussed by Ruth Kastner in a blog post~\cite{Kastner2018}.} In isolation, this passage might seem make locality unfalsifiable. Of course experiences only happen to an agent along her worldline---something like this would be true in any universe that contains embodied agents. What would stop us from telling the same story about a hypothetical physical theory that admitted the transmission of information faster than the speed of light? This can't be the only reason QBism is local, otherwise QBism could never be nonlocal if ever called on by a future physics to be so.

In the next section, we begin to make amends for this deficit. In fact, we believe most QBists have had the proper intuition on this point all along, making it curious that it has not appeared in print.\footnote{Examination of email records reveals that, as early as February 2013, Schack had produced drafts towards a paper that was intended to articulate most of the points we make here. Although this discussion inexplicably never saw the light of day in print, a recorded lecture from 2015 \cite{Schack2015} reveals the QBists had, at least internally, resolved the issue to their satisfaction.}  More importantly, however, many of QBism's commentators, critics, and even sympathizers have not understood the proper sense in which QBism can be said to be local. There may be other aspects of QBism which are legitimately challengeable, but in this paper we hope to convince the reader that its claim to locality is not among them.

\section{Why QBism is local}
\label{AllThreeOfThem}

Once the imagery of a shared spacetime arena is shed, it becomes readily clear where QBism must locate its locality: in the judgments of an individual. But how can this be? `Over there' is supposed to be a place, not a belief. Here it is important to remember that QBism considers quantum theory to be an organizing structure for detecting inconsistencies in one's mesh of beliefs. This means one brings a whole life of expectations \emph{to} quantum theory whenever one makes use of it.  Expectations are the inputs, not the outputs. The theory's outputs are whether or not the expectations are coherent with each other and with \emph{the character of the world}~\cite{DeBrota2021,DeBrota2024}.  The role of quantum theory is not to tell an agent what they must believe because of the initial conditions of the universe or some stand-in for the same. Nor does quantum theory have the power within it to demand that disparate agents must come to agreement.  Quantum theory in QBism just isn't a direct picture of what is `happening' in the world.

It is worth noting that this is essentially the antithesis of what Adlam~\cite{Adlam2022}, a frequent critic of QBism, demands of quantum theory. In the worldview she seems to espouse, quantum theory is (meant to be)\footnote{We say `is meant to be' because Adlam thinks it likely that quantum theory will need to be augmented by a deeper physics in order to get what she wants~\cite{Adlam2026}.} a direct description of what is happening in the world, and so things like agents and decision makers and the expectations they hold must be derivable from within the theory itself: They must be among the theory's outputs. To use this as a criticism of QBism is to simply deny what QBism and its research program are about.

In QBism, quantum theory does not give materiality to space and time any more than it does to atoms and chairs. The QBist approach legitimizes talk of the `now' in science~\cite{Mermin2013,Mermin2014} because it considers quantum theory to be a tool we bring to our experiences rather than something that tells us what our experiences will be. What is `here' and what is `there' are primitives for the particular user and refer to her expectations for the consequences of certain actions.

What is `here' for a particular agent is something she believes she can act upon immediately. She calls something `far away' if she believes she cannot take a direct action on it, but must take intermediate actions on other things `in between' first. Notice that she can use the quantum formalism to account for both the close and the far. In the EPR scenario, Alice assigns an entangled state to the particles. This assignment reflects her expectation that one particle is at her detector, ready for her to measure, and that the other is at Bob's detector, which she can only access if she makes a trip to it. Her state is an operator on the joint Hilbert space $\mathcal{H}_A\otimes\mathcal{H}_B$, but, at present, she can only act on the $\mathcal{H}_A$ subsystem. 

In this view, a spacetime diagram is a sophisticated way for an individual to account for experiences she has already had along with the ones she is still to have~\cite{Mermin2013,Mermin2014}. As David Mermin put it,
\begin{quote}
[A spacetime diagram] enables me to represent events from my past experience, together
with my possible conjectures, deductions, or expectations for events that are not in my
past, or that escaped my direct attention. By identifying my abstract diagram with an
objective reality, I fool myself into regarding the diagram as a 4-dimensional arena in which
my life is lived.

The events I experience are complex extended entities and the clocks I use to assign
times to my experiences are extended macroscopic devices. To represent my actual experiences as a collection of mathematical points in a continuous space-time is a brilliant
strategic simplification, but we ought not to confuse a cartoon that concisely attempts to
represent our experience, with the experience itself. (\cite[p.\ 4]{Mermin2013})
\end{quote}
In fact, insofar as both are extracted from an agent's personal judgments, a spacetime diagram and a quantum state assignment have a similar ontological status in QBism. Within the diagram, an agent's possible experiences are all in her future light-cone. She may also record her beliefs about previous experiences as taking place along a worldline terminating at the present moment, which, from each past instant to the next, traces out a path within the light-cone of each spacetime point. In this way, the diagram codifies pictorially a structural aspect of her beliefs, namely, that her current actions can only have consequences in her causal future. Bob's subsystem is of relevance to Alice because it will eventually enter her experience, either directly when she makes a trip to it or through an appropriate surrogate such as a phone call with Bob (thus, taking an action on Bob). The space-like separation Alice attributes to Bob's action at the beginning of the experiment captures her belief that her future encounter with either Bob or the system will be independent of the choice she makes now about what to measure on the system in front of her. 

To build a picture of how QBism thinks drastically differently from other interpretations of quantum theory, let us first describe the setting of Bell's argument from the usual point of view. There, one imagines Alice and Bob with an infinite supply of two spatially separated quantum systems, each pair prepared in identical entangled quantum states. The disparate systems are considered to be localized in front of Alice and Bob so that they may make simultaneous measurements on them as defined by a common frame of reference---Alice and Bob's rest frame. The usual story goes that Alice chooses some measurement $A$ (with outcomes $a$) and Bob chooses some measurement $B$ (with outcomes $b$), each of their own free will with each entangled pair. Their separate detectors go ``click'' in each instance, giving settings and outcomes that they can record in their separate sites to compare later for ferreting out any nefarious correlations.

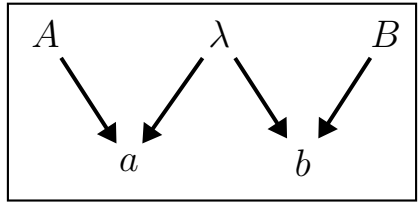
\begin{figure}[ht]
\centering
\begin{tikzpicture}[>=Triangle, thick, every node/.style={font=\Large}]
\draw[line width=0.8pt] (-0.5,-0.5) rectangle (4.9,2.1);
\node (A) at (0,1.7) {$A$};
\node (lambda) at (2.3,1.7) {$\lambda$};
\node (B) at (4.5,1.7) {$B$};
\node (a) at (1.1,0) {$a$};
\node (b) at (3.4,0) {$b$};
\draw[->, line width=1.5pt] (A) -- (a);
\draw[->, line width=1.5pt] (lambda) -- (a);
\draw[->, line width=1.5pt] (lambda) -- (b);
\draw[->, line width=1.5pt] (B) -- (b);
\end{tikzpicture}
\caption{Bell's assumptions illustrated from a third-person perspective.}\label{fig:GodBell}
\end{figure}

Notice that all of this is presented from a kind of God's eye view and that the very availability of this view is the content of Bell's `realism' assumption. All variables in the scenario are conceptually marks on a single register, which can be viewed in its totality and all at once. The values of these variables, moreover, are taken to be governed by objective probabilities imagined to be entailed by the experimental arrangement. If we let $\lambda$ further symbolize the initial preparation, Bell's assumptions can be summarized in the causal diagram depicted in Figure \ref{fig:GodBell}. Arrows in such a diagram indicate statistical dependence, while the lack of an arrow implies statistical independence. Hence, because of `locality', $a$ depends on Alice's setting $A$ and the state of the system $\lambda$, but not on Bob's setting $B$ or outcome $b$. The settings $A$ and $B$ similarly do not depend on anything from their other sides. Finally, because of `measurement independence', there are no arrows between the settings $A$ and $B$ and the shared source $\lambda$. It is well-known that these probabilistic constraints imply the Bell inequalities that quantum mechanics violates, requiring the relaxation of at least one of Bell's assumptions.

Here is the crucial point. If a third-person, God's eye view, or view from nowhere description of the whole scenario, with preparation, choice of settings, and outcomes taking place in a single shared spacetime arena, were simply logically necessary or \emph{analytic}---made true by the very meaning of the words employed in the thought experiment---then, indeed, rejecting `realism' would not actually be a live option, as some philosophers of science have claimed. One would indeed need to choose between rejecting `locality' or `measurement independence'. But inevitable as it might seem to certain temperaments, such a description is both formally and practically unnecessary.\footnote{While QBism is far from alone in observing that `realism' or AOE may be coherently rejected, this is where most of its similarities with other interpretations ends. In a recent taxonomy \cite{DeBrota2026}, QBism formally stands out when the `realism' escape route is disaggregated into four distinct theses that may be individually rejected.} So how does QBism, after all, think differently?

The key difference in QBism is that all of this must be presented from a first-person point of view---for that is the very meaning of seeing quantum theory as a decision theory. QBism does not assume that we are entitled to take up a view from nowhere.\footnote{This treatment originally arose around QBism's commitment to personalist Bayesian probabilities, which powered the rejection of the notion that there was any such thing as \emph{the} probability distribution governing the random variables in the Bell scenario. All probabilities must instead be personal probabilities, tied to a particular agent. A full commitment to this thesis brought to the particularities of quantum theory led to the full-blown metaphysical notion that there is no view from nowhere.} Consequently, the Bell scenario must be told from a particular situated perspective. We could take the perspective of Alice, or of Bob, but neither participant can simultaneously ``see'' the settings and outcomes on both sides\footnote{Of course, one might ask, ``Why not have a super-observer Charlie overseeing it all?'' Surprisingly, maybe the sharpest response to this can be borrowed from a committed Everettian, David Wallace~\cite[p.\ 310]{Wallace2012}: ``Bell’s theorem \ldots\  assumes, tacitly, among its premises that experiments have unique, definite outcomes. \ldots\ From the perspective of a given experimenter, of course, her experiment does have a unique, definite outcome, even in the Everett interpretation. But Bell’s theorem requires more: it requires that from her perspective, her distant colleague’s experiment also has a definite outcome. This is not the case in Everettian quantum mechanics \ldots.'' Nor is it the case in QBism: When Alice is using quantum theory, it is solely for her ascribing probabilities to her own experiences. As Fuchs, Mermin, and Schack \cite{Fuchs2014} put it, ``At the moment of his own measurement Bob is playing the friend to Alice’s far-away Wigner.''  Of course, Wallace goes on to say, ``And from the third-person perspective from which Bell’s theorem is normally discussed, no experiment has any unique definite outcome at all.'' To which the QBist street punk responds, 
``QBism don't do third-person.''}---to be able to do so would be to say that the experimenters and their equipment were not, after all, `far away'. In what follows, we will take Alice's perspective. 

As we are specifically concerned with Alice's experiences, we need to be more precise about our variable definitions: In QBism's account, $B$ and $b$ cannot be treated as variables with a stand-alone existence at Bob's site.  They can only denote consequent experiences for Alice elicited by taking actions on ``Bob's measurement device'' or ``measurement record.'' In effect, Alice takes the required physical action to ask of Bob, ``What did you do, and what did you find?''  We might call $B$ and $b$ ``Bob's measurement choice'' and ``Bob's outcome,'' respectively, but in QBism, they are \emph{originarily} experiences of Alice alone, the decision maker using the quantum formalism.\footnote{In phenomenology, especially Husserl's, originarily (origin\"ar) refers to what is given directly and primitively, prior to theoretical interpretation or conceptual mediation.}

The measurement setting $A$ is Alice's freely-chosen action, but the variables $a$, $B$, and $b$ are thus events for her. In particular, both $B$ and $b$ are necessarily events in Alice's future. Finally, $\lambda$ symbolizes Alice's relevant beliefs about the system in virtue of which she assigns her initial quantum state. On the basis of the experimental setup and her belief in relativistic locality, Alice's probabilities satisfy certain structural constraints. First, she believes that Bob's measurement choice does not depend on the state of the system, that is, $P(B|\lambda)=P(B)$. And second, because of the space-like separation, she believes that neither Bob's setting nor his outcome depends on her measurement choice, that is, $P(B|A)=P(B)$ and $P(b|A)=P(b)$. 

It is crucial to realize that while locality implies Alice believes $b$ is independent of her measurement setting $A$, it \emph{does not} imply that Alice should believe that $b$ is independent of her \emph{outcome} $a$. Indeed, generically, $P(b|a)\neq P(b)$. Why is this so? Because Alice learns something when she experiences $a$. This is simply probabilistic updating. If Alice begins with a maximally-entangled quantum state and measures her particle, it is a feature of her updated state that, if she learns the measurement basis on Bob's side is the same as hers (i.e., by performing the appropriate measurement to find out), she will become \emph{certain} of its outcome before performing the necessary measurement to elicit it as well. Yet, as Fuchs~\cite{Fuchs2002} and DeBrota and Stacey~\cite{DeBrota2018FAQBism} emphasized before, there is nothing nonlocal about this---it simply reflects that there are correlations in Alice's probabilities for the two subsystems.

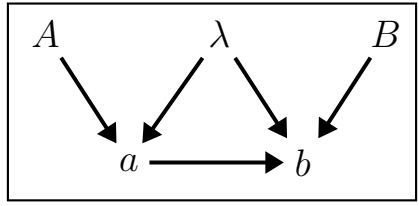
\begin{figure}[ht]
\centering
\begin{tikzpicture}[>=Triangle, thick, every node/.style={font=\Large}]
\draw[line width=0.8pt] (-0.5,-0.5) rectangle (4.9,2.1);
\node (A) at (0,1.7) {$A$};
\node (lambda) at (2.3,1.7) {$\lambda$};
\node (B) at (4.5,1.7) {$B$};
\node (a) at (1.1,0) {$a$};
\node (b) at (3.4,0) {$b$};
\draw[->, line width=1.5pt] (A) -- (a);
\draw[->, line width=1.5pt] (lambda) -- (a);
\draw[->, line width=1.5pt] (lambda) -- (b);
\draw[->, line width=1.5pt] (B) -- (b);
\draw[->, line width=1.5pt] (a) -- (b);
\end{tikzpicture}
\caption{The probabilistic constraints on Alice's beliefs as they arise in QBism under the assumption of locality.}
\label{fig:AliceBell}
\end{figure}

Putting these constraints together into a causal diagram yields Figure \ref{fig:AliceBell}. Notice that it is the same as Figure \ref{fig:GodBell} except for the additional dependence of $b$ on $a$. When we refrain from imagining we can see through God's eyes, and instead put ourselves in Alice's shoes, we end up with a different causal diagram. It should now be clear how QBism avoids deriving Bell inequalities: With the additional arrow in Figure \ref{fig:AliceBell}, these constraints do not entail, for instance, a Clauser-Horne-Shimony-Holt (CHSH) \cite{CHSH1969} bound of $2$ for quantum mechanics to violate. Following Jarrett~\cite{Jarrett1984}, in the terminology of Shimony~\cite{Shimony1986}, these constraints violate outcome independence (OI), but not parameter independence (PI). When `realism' is assumed and we have objective probabilities, Bell's `locality' is the conjunction of OI and PI. However, with the personal probabilities and outcomes of QBism, OI is not about locality at all.\footnote{Starting from other considerations, a notion of locality that does not assume OI was used by Colbeck and Renner in~\cite{Colbeck2011}.} As we saw above, OI is violated by probabilistic updating. If probabilities are personal, OI simply doesn't make sense. Instead, in QBism, PI is the meaningful notion of locality that remains once hidden variables are rejected in this way.\footnote{PI alone implies only the linear constraints that define the no-signaling polytope, which, in the Bell scenario, coincides with the trivial bound of 4 for the CHSH quantity~\cite{Popescu1994}.} Instead of sacrificing locality for the sake of hidden variables, QBism explicitly keeps locality precisely by rejecting hidden variables. Bell's `realism' demands a single objective probability function, not belonging to any of the participants, and reflecting the assumption that all measurement events from both Alice and Bob are taking place within one spacetime arena. But in QBism, probabilities and spacetimes alike are personal. Just as there is no view from nowhere, there are no probabilities from nowhere, either.

\begin{figure}[h]
\includegraphics[width=\textwidth]{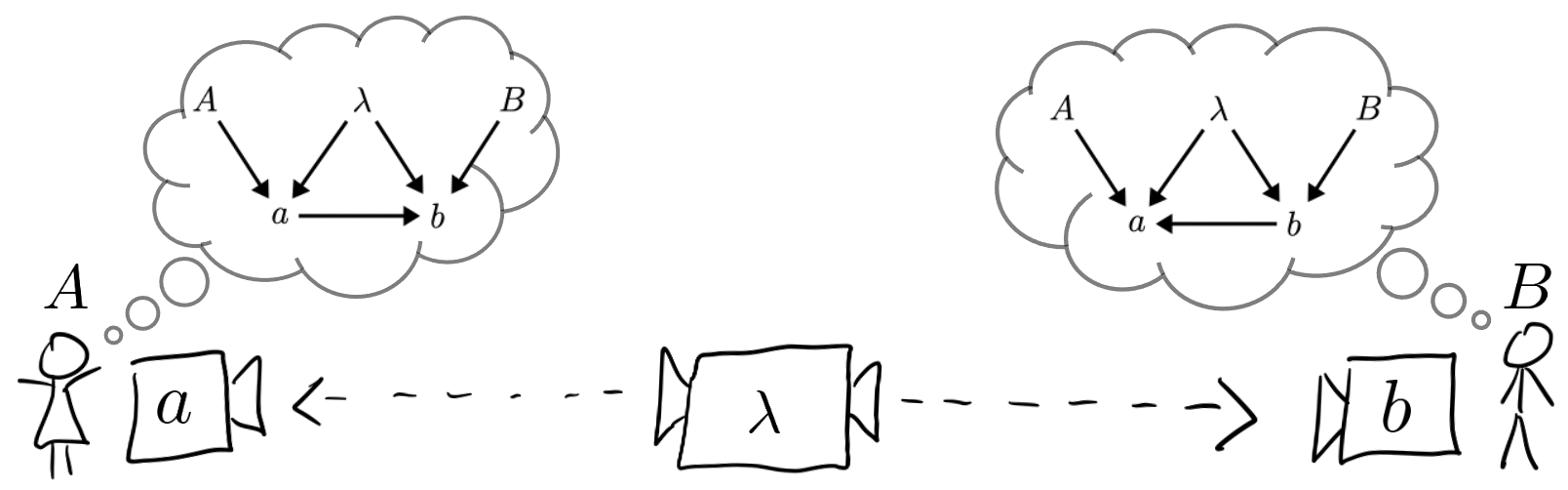}
\caption{The full experimental scenario with each agents' personal causal diagram.}\label{fig:AliceBobBell}
\centering
\end{figure}

What about Bob? In principle, he's a completely different story with completely different priors---he is, after all, a different person. But if we assume he and Alice are sufficiently on the same page to assign the same joint state and measurement operators, then everything goes through as before, but with the appropriate reinterpretations of $A$ and $a$ as experiences for Bob. For the same reason as before, there will be a dependence between $a$ and $b$, but now pointing in the other direction. The overall situation is illustrated in Figure \ref{fig:AliceBobBell}.

When we ask whether quantum mechanics is local, we are asking whether the theory respects the notion of locality that we inherit from relativity. The locality in question is about whether or not a single agent's beliefs, reflected in her probability assignments, can accommodate any quantum scenario without violating the constraints of locality. And now the answer is clear: When we bring quantum mechanics to our standard expectations of relativistic locality, constraining all processes to be properly local, nothing in the formalism, entanglement or otherwise, requires any revision to this assumption. Quantum theory does not force us to abandon our belief in locality to admit influences between the systems we designate as space-like separated. It is in this sense that QBism is local.

\section{What would a nonlocal QBism look like?}

However, in addition to not forcing a revision of our beliefs in locality, quantum theory doesn't force adherence to them either. Not being forced to abandon one's relativistically local beliefs is the pertinent notion of locality in, for instance, Bell's theorem, and so we prefer this terminology. One could also make the argument for a term like ``causally neutral'' to indicate that quantum theory can be brought to any causal structure.\footnote{By now it should be clear that for us a causal structure is just a summary of the restrictions an agent adopts for her probability assignments.} In a slogan, ``The quantum formalism doesn't know anything about space and time.''

This fact suggests there are two ways genuine nonlocality could crop up such that even a QBist would take note. The first way is if a future revision to quantum theory actually failed to be causally neutral. This would mean that, rather than a reasoning structure we bring to our prior beliefs about space and time, this future theory explicitly violates relativistic locality. Even though quantum theory would be a particular kind of effective approximation to it, this theory would force a user to admit that her beliefs must allow for influences between regions that she deems to be space-like separated. It is not impossible that a future theory of quantum gravity could take such a form. While a broadly QBist reading could probably still be given to a theory of this sort, it would remain to be seen whether or not the specifics of the theory undermine the original arguments for QBism. For example, if in this future theory it turns out that quantum states can, after all, be cloned, the grounds for considering a quantum state as someone's personal judgment would evaporate. More generally, the reduced flexibility implicit in no longer being causally neutral might make this future theory seem less like a means for detecting inconsistencies among our beliefs and more like one that tells us what our beliefs should be, or one that simply mirrors what is happening in the world.

The second way for a proper nonlocality to appear involves no modification to the quantum formalism at all. This could turn up, for instance, if one comes to believe that a process is best accounted for in the quantum formalism by introducing a nonlocal operation---for instance, if it had turned out to be fruitful to quantize the Wheeler-Feynman electrodynamics mentioned in the introduction~\cite{Feynman1966}.  Or relatedly, if a QBist found it easiest to explain their experimental results with a Hamiltonian with interaction terms that couple regions that are space-like separated. This would explicitly lead to what is equivalent to action at a distance within that particular agent's expectations (and yet, of course, their experiences would still come to them along their own worldlines). If they think such a process is available, they will expect that the probabilities of future measurement outcomes on a far away system could fail to be independent of their freely-chosen actions on a system here. While the quantum formalism itself does not forbid writing down such nonlocal Hamiltonians, it is so far thought to be an empirical fact that all natural processes can be adequately accounted for with only interaction terms that respect the usual spacetime causal structure. The main point is, an agent's causal structure is an issue unrelated to the quantum formalism. This has practical implications as it means that revisions to quantum theory and revisions to relativity, if ever needed, could be considered independently.

Two remarks are in order. First, as we explained in the previous section, but can never emphasize enough, entanglement has nothing to do with nonlocality in the QBist story. The entangled correlations in the EPR scenario have a local origin and nowhere does anyone employ nonlocal operations. Quantum correlations just aren't spooky in that way. Second, it should be clear that the standards that must be met for actually accepting evidence of any kind of nonlocality would and should be monumental. It would be truly astonishing that all known physics had perfectly respected locality in the pertinent way only for a genuine violation to fly under our noses all this time. While possible, until independently verified again and again, it would be far more reasonable to assume that a wire was loose somewhere. 

So, although we believe the prospects for either kind of nonlocality are dim, we do not go so far as ruling out the possibility on logical grounds. Moreover, if nature were nonlocal, the case for a QBist reading of quantum theory would certainly be weakened---that's the way science works. Yet quantum theory, by itself, does not push us in this direction, and QBists believe the lessons of quantum theory for reality are of a remarkably different flavor. 

\section{What is reality like if not nonlocal?}

Hopefully it is now clear what QBists mean when they say their interpretation is local. It makes sense, at least in their terms, to use this terminology. Now, one might demand that the term `locality' should mean more than the QBists are willing to grant it, perhaps by demanding that every measurement event, from every observer, take a unique value in a single spacetime. But to demand this is to build-in an acceptance of AOE. This move, of course, severely limits the options for a denial of nonlocality, leaving, for example, some interpretations which dispense with measurement independence. Even someone strongly committed to AOE should, however, recognize that it is not mandatory and that relaxations of it can remain properly local.

The `fragmentation' of spacetime announced by the title of Pienaar's 2026 article \cite{Pienaar2026} points toward the kind of ontological lesson a QBist might draw. As Pienaar notes, interpretations like QBism appear to be incompatible with the so-called `block universe'. Indeed, QBism looks for an ontology which does not consider everyone to be an object within one big arena. This imagery might approximately work in certain circumstances, but it cannot withstand the strain of others and breaks completely in Wigner's friend scenarios. It simply doesn't work, given QBism's other commitments. 

What might an alternative look like? At present, we only have hints to work with, but the absence of a robust proposal does not weaken the argument that an alternative is needed---after all, if we knew the solution already, there would be nothing to research. In a recent paper \cite{DeBrota2024}, greatly building on an old idea \cite{Fuchs2012b},\footnote{Further evidence that, as this paper illustrates, QBism moves in rather slow, but methodical ways.} we showed how quantum dynamics can be derived from consistent reasoning about possible measurements, more fully establishing the long-held QBist dictum that quantum theory concerns measurement actions alone, and that it is in the moments of creation implicit in each such act that we might hope to find an appropriate metaphysics. One can imagine, then, that the structure of quantum measurement supplemented by structural elements of individuals' beliefs, such as the belief in locality, may help to furnish a suitable replacement for the block.

From a different kind of angle, DeBrota and List \cite{DeBrota2026} sketch two potential refinements of the ontological position QBism may be read as taking. In the resulting taxonomy, QBism rejects either the `one world' thesis or the `non-fragmentation' thesis. They nickname the former option Pluriverse QBism and associate a subjective or personal `world' to each agent. This move would align with Pienaar's suggestion: 
\begin{quote} 
``One strategy which appears natural in light of the present work would be to define `locality' as the requirement that relative locality holds for each observer (as defined in the previous section). Accordingly, spacetime concepts should always be indexed to a specific observer, for instance we may take the ``spacetime of observer $A$'' to refer to the manifold $\mathcal{M}_A$ equipped with some metric, such that spacetime intervals are only defined between events that occur relative to $A$ and which are therefore embeddable in $\mathcal{M}_A$.'' (\cite[p.\ 15]{Pienaar2026}) 
\end{quote} 
The other option, nicknamed Fragmentalist QBism, preserves the notion of there being one world at the cost of this world formally admitting incoherence. There could be coherent `fragments' in the world, yet `all that is the case' will not fit together into one coherent whole, somewhat reminiscent of an impossible Escher painting. It remains to be seen how this kind of picture could be substantiated technically.

More broadly, we suspect that it is, at some level, the unavailability of a block picture that has made the reconciliation of quantum mechanics and gravity so elusive. It might even be that working out a successful alternative in the nonrelativistic setting will be necessary for imagining the right kind of generalization. For us, it is a fascinating and exciting possibility that we might reject a lifeless block reality; that we might, perhaps, embrace a reality with a constitutive role for lived experience, that is, like our active perceptual engagement, continuously unfolding, making and remaking itself, and never fully determinate---``await[ing] part of its complexion from the future'', as William James \cite{James1907} wrote. That we might be participants in a reality where all structure emerges from a primordial lawlessness, as John Wheeler \cite{Wheeler1983} once dreamed. And yet, with all these radical departures from tradition on the table, all current uses of quantum theory in QBism remain local.

\printbibliography
\end{document}